\documentclass{aa}
\usepackage{savesym}
\usepackage{graphicx}
\usepackage{natbib}
\usepackage{amssymb}
\usepackage{amsmath}
\savesymbol{iint}
\usepackage{txfonts}
\restoresymbol{TXF}{iint}

\titlerunning{IR absorption of helium in cool white dwarfs}
\authorrunning{Kowalski}

\offprints{Piotr Kowalski, \email{p.kowalski@fz-juelich.de
}}

\begin{document}

\title{Infrared absorption of dense helium and its importance in the atmospheres of cool white dwarfs\footnote{The opacity table is available in electronic form
at the CDS via anonymous ftp to cdsarc.u-strasbg.fr (130.79.128.5) or via http://cdsweb.u-strasbg.fr/cgi-bin/qcat?J/A+A/
}}
\author{Piotr M. Kowalski}

\institute{IEK-6 Institute of Energy and Climate Research, Forschungszentrum J\"{u}lich, Wilhelm-Johnen-Strasse, 52425 J\"{u}lich, Germany}

\abstract{}
{
Hydrogen deficient white dwarfs are characterized by very dense, fluid-like atmospheres
of complex physics and chemistry that are still poorly understood. The incomplete description 
of these atmospheres by the models results in serious problems with the description of spectra of 
these stars and subsequent difficulties in derivation of their surface parameters. Here, we address the problem 
of infrared (IR) opacities in the atmospheres of cool white dwarfs by direct $ab$ $initio$ simulations of IR absorption of dense helium.
}
{
We applied state-of-the-art density functional theory-based quantum molecular dynamics simulations to obtain
the time evolution of the induced dipole moment. The IR absorption coefficients were obtained by 
the Fourier transform of the dipole moment time autocorrelation function.
}
{
We found that a dipole moment is induced due to three- and more-body simultaneous collisions 
between helium atoms in highly compressed helium.
This results in a significant IR absorption that is directly proportional 
to $\rm \rho_{\rm He}^3$, where $\rho_{\rm He}$ is the density of helium. To our knowledge, this absorption mechanism 
has never been measured or computed before and is therefore not accounted for in the current 
atmosphere models. It should dominate the other collisionally induced absorptions (CIA),
arising from $\rm H-He$ and $\rm H_2-He$ pair collisions, and therefore shape the IR spectra of helium-dominated 
and pure helium atmosphere cool white dwarfs for $\rm He/H>10^4$.
}
{
Our work shows that there exists an unaccounted IR absorption mechanism arising from the multi-collisions between 
He atoms in the helium-rich atmospheres of cool white dwarfs, including pure helium atmospheres. This absorption
may be responsible for a yet unexplained frequency dependence of near- and mid- IR spectra of helium-rich stars. 
}

%\end{abstract}


\maketitle

\section{Introduction}

\begin{figure}
\resizebox{\hsize}{!}{{\includegraphics{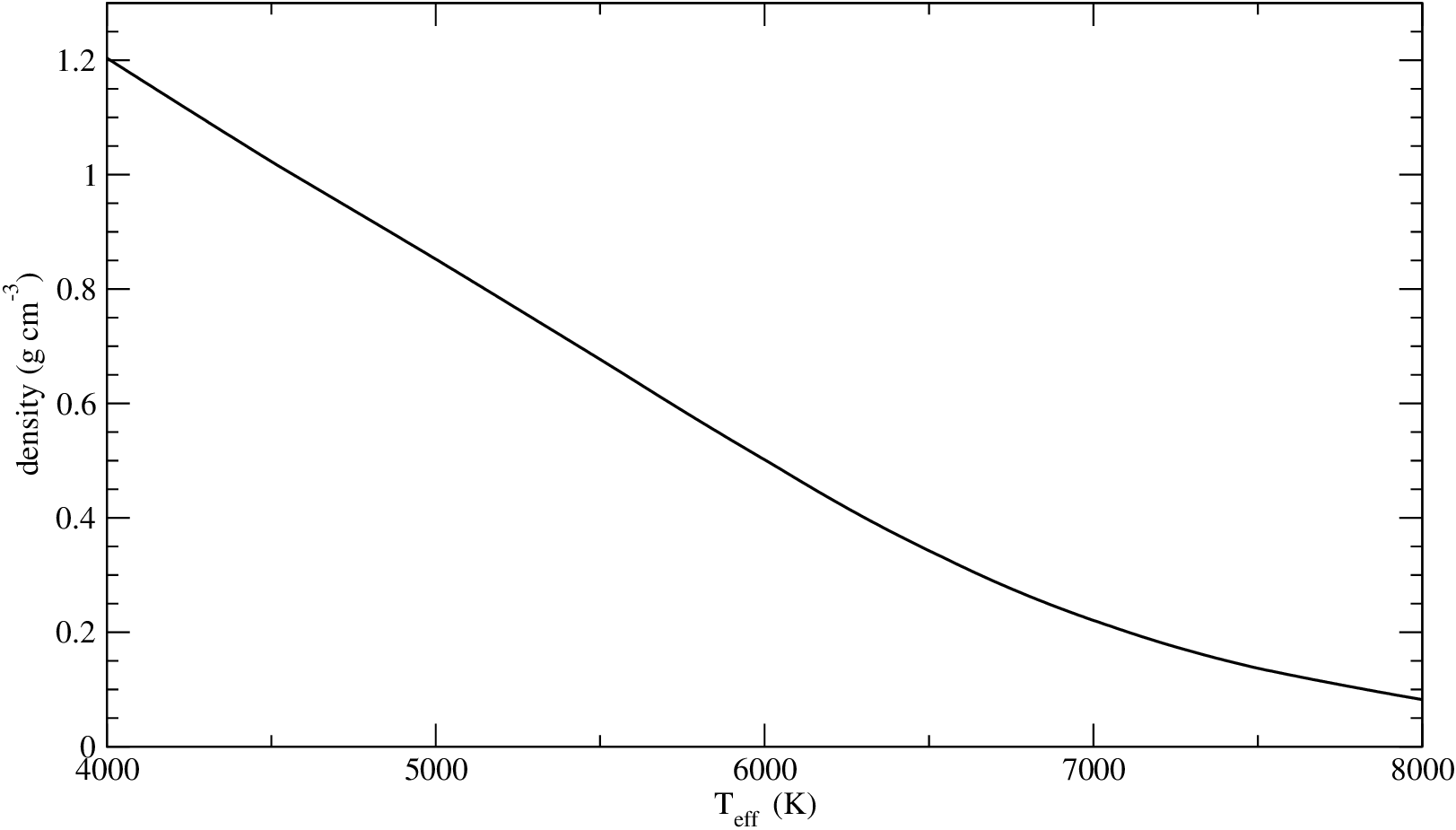}}}
\caption{The photospheric density ($\tau_r=2/3$) in the pure He white dwarf atmospheres of given 
$T_{\rm eff}$, as predicted by our atmosphere models (see the text for models details).
\label{F1}}
\end{figure}

Being billions of years old cool white dwarfs with $T_{\rm eff}\rm <8000\,K$ have received significant 
attention because they can be used as cosmochronometers \citep{LL84,FN00}.
Their spectra carry important information that, when correctly decoded, can tell us about physical parameters, such as $T_{\rm eff}$, gravity, 
and chemical composition of their atmospheres, which reveals in return information about past stellar and planetary formation processes 
prevailing in our Galaxy \citep{F09,FN00,RI06}. This information could be correctly deciphered only when a reliable set of the atmosphere 
models is used in the analysis \citep{KSH13,K07}. Unfortunately, having fluid-like densities of up to a few $\rm g/cm^3$ (see figure \ref{F1}),
the atmospheres of helium-rich white dwarfs are very difficult to model. Over the last decade, several dense-fluid effects were introduced into the modeling
that substantially improved the description and understanding of the atmospheres of these stars. These include the pressure-induced absorption mechanisms 
\citep{IRS02,KS06,K06b,K10,GF01}, the refractive radiative transfer equation \citep{KS04}, the non-ideal equation of state and chemistry of spectroscopically 
important species \citep{KSM07,K06a} and also the recently improved $\rm H_2$-He collision induced absorption \citep{AF12}.
However, because of their complex physics and chemistry, the atmospheres of helium-rich, cool white dwarfs 
are still poorly understood, which is indicated by poor fits to the spectral energy distributions of these stars by models \citep{BL02,K09,KSH13}.
Most of these problematic stars, such as LHS3250 \citep{HD99,BL02}, or the several so-called ultra-cool white dwarfs that were discovered thanks 
to the Sloan Digital Sky Survey \citep{G04,BL02} show significant near- and mid- IR flux depletion that is thought to be caused
by the strong $\rm H_2-He$ and $\rm H-He$ CIA absorptions in extremely dense, helium-dominated atmospheres ($\rm He/H>>10^{3}$).
However, none of such spectra, especially their near- and mid- IR parts, could be successfully fitted by the current models.

In this contribution, we address the problem of IR absorption by performing state-of-the-art {\it ab initio} molecular dynamics simulations 
of the IR opacities of dense helium. Because IR opacities from pure helium have never been reported before, we were looking for any IR absorption signatures caused by fluctuations 
of the dipole moments induced in highly compressed helium. We were especially interested in its strength and importance in the atmospheres
of helium-rich, cool white dwarfs. Our {\it ab initio} simulations have revealed the then unknown infrared absorption mechanism that shapes 
the IR spectra of helium-rich atmosphere, cool white dwarfs. We show that this absorption, which is a few orders of magnitude weaker 
in absorption cross-section than already accounted for in the models $\rm H-He$ and $\rm H_2-He$ CIA opacities, dominates the IR absorption when $\rm He/H>10^{4}$,
including the pure He case.

\section{Computational approach}

Our simulations were performed by application of density functional theory (DFT) method, which is widely used in the quantum mechanical computations of dense, 
complex, many particle systems \citep{KH00}. The DFT methods are appropriate for calculations of the ground state energies of multi-electron 
systems, and have proven to be very useful in calculations of various chemical and physical properties of atoms, molecules, solids, and matter under high compression   \citep{JK14}, including those
constituting the atmospheres of cool white dwarfs (e.g., \citet{KSM07,K10}). In our work, we used one of the most common implementations of DFT, the generalized gradient 
approximation (GGA) with PBE exchange-correlation functional \citep{PBE96}. We used plane-wave DFT CPMD code \citep{CPMD} with ultrasoft pseudopotentials 
\citep{DV90}, and the energy cutoff of $\rm 340 \, eV$. The Born-Oppenheimer molecular dynamics simulations were performed on 32 atoms containing 
supercells, and in each simulation, a $160\rm \,ps$ long trajectory was generated. The total dipole moment of the supercell was computed in every step
with the timestep of $1.2\,\rm fs$. We notice that we neglect the higher frequency simulation results in the analysis because this timestep gives good sampling of frequencies up to about $6000\,\rm cm^{-1}$.

\begin{figure}
\resizebox{\hsize}{!}{\includegraphics{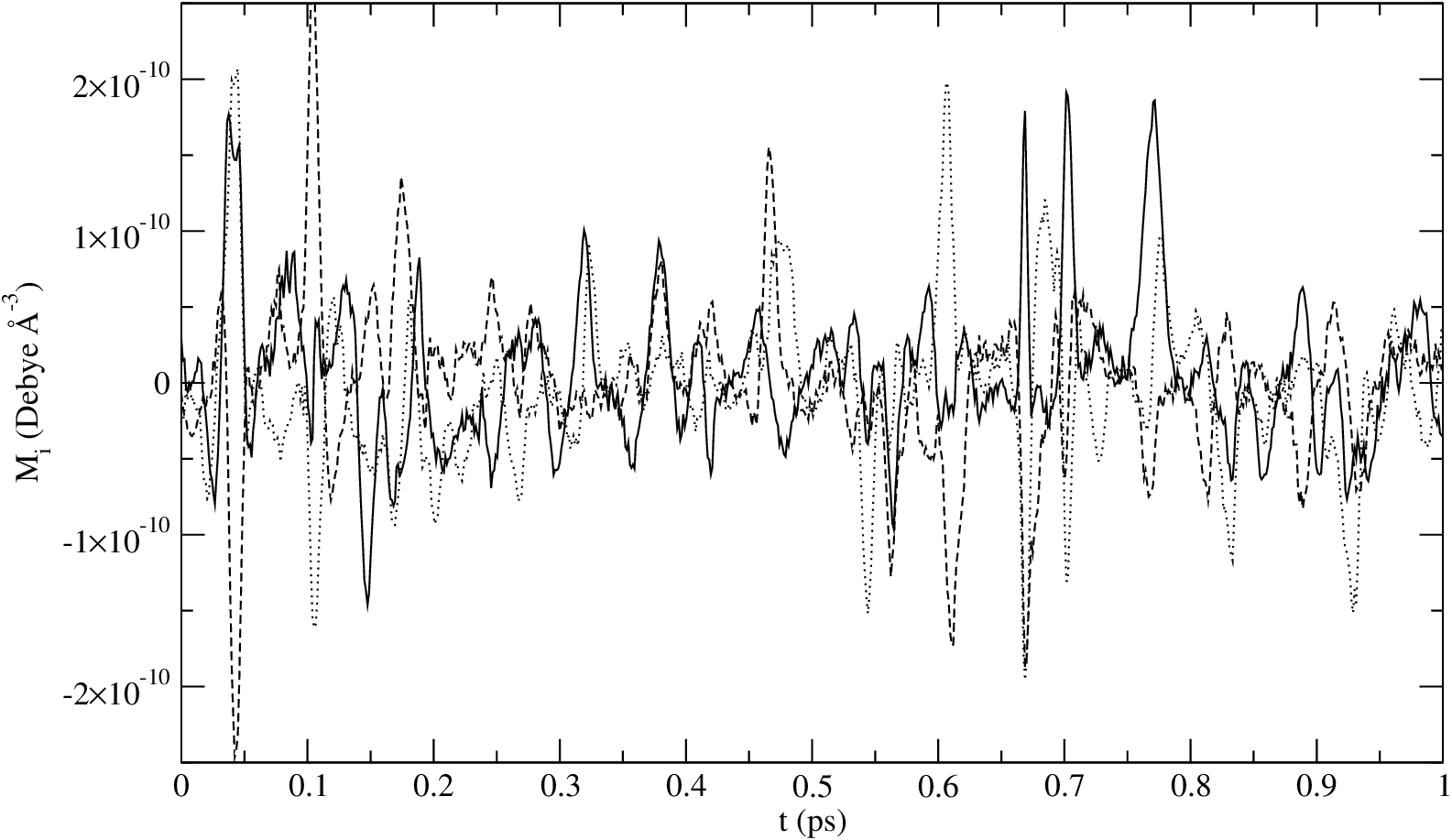}}
\caption{ The time evolution of the components of the dipole moment vector $\bf M$ in dense helium of
$T=\rm5000\,K$ and $\rho_{\rm He}=514\,\rm amagat$, where amagat=$2.68678\cdot 10^{19}\,\rm cm^{-3}$. 
The lines mark $M_x$ (solid), $M_y$ (dotted), and $M_z$ (dashed) components.
\label{F2}}
\end{figure}

The IR spectrum is represented by the frequency-dependent absorptivity coefficient, $\alpha (\omega)$, that was computed through 
the Fourier transform of the dipole moment time autocorrelation function \citep{G91,SBP97,JK14} as
\begin{equation}
%\begin{split}
\alpha(\omega)=\frac{2\pi \omega^2}{3 c k_{\rm B}T V}
\int_{-\infty}^\infty dt\exp(-i\omega t)\left< {\bf M}(t) \cdot {\bf M}(0)
\right>,
%\end{split}
\end{equation}
where $c$, $K_{\rm B}$ and $V$ are the speed of light, the Boltzmann constant, and the supercell volume respectively. 
The vector ${\bf M}(t)$ is the total dipole moment of the simulation cell at a time $t$. The method of computing IR opacities by the means 
of molecular dynamics simulations has been used in the past, for instance for calculation of IR spectrum of water \citep{SBP97,IT05,G91}.
We note that the real absorption coefficient that is used in the modeling
is given by $\alpha(\omega)/n(\omega)$, where $n(\omega)$ is the index of refraction, because an atmosphere of a cool, helium-rich white dwarf is a refractive medium \citep{KS04}. 
In Figure \ref{F2}, we show an example of the time evolution of the ${\bf M}(t)$ vector components that gives rise to the non-negligible IR absorption by dense helium.

To compute the atmosphere models we used our own stellar atmosphere code that accounts for various dense medium effects, like the refraction \citep{KS04}, 
the non-ideal equation of state and chemical equilibrium \citep{K06a,KSM07} and the high density corrections to the important absorption mechanisms \citep{IRS02,KS06,K06b}. 
It also includes the recently improved $\rm H_2-He$ CIA opacities of \citet{AF12}.

\begin{figure}
\resizebox{\hsize}{!}{{\includegraphics{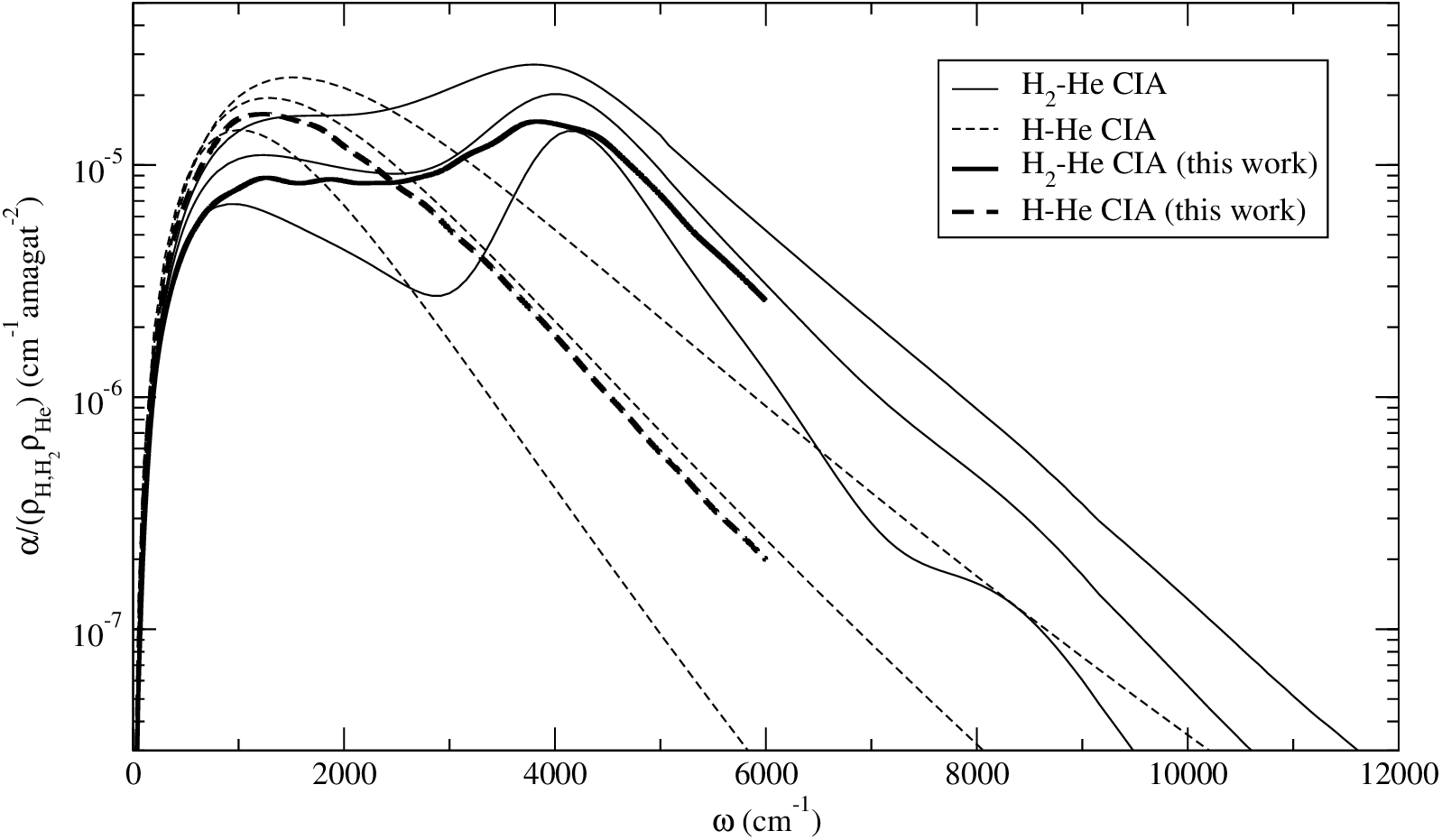}}}
\caption{
The $\rm H_2-He$ (thin solid lines, \citet{AF12}) and $\rm H-He$ (thin dashed lines, \citet{GF01}) CIA opacities. 
The results for $T\rm =7000\,K$, $\rm 5000\, K$ and $\rm 3000 \, K$ are presented (from top to bottom).
The thick lines represent the result of our simulation performed for $T=\rm5000\,K$, $\rho_{\rm H_2,H}=16\,\rm amagat$, and $\rho_{\rm He}=498\,\rm amagat$, where 
amagat=$2.68678\cdot 10^{19}\,\rm cm^{-3}$.
\label{F3}}
\end{figure}

\begin{figure}
\resizebox{\hsize}{!}{{\includegraphics{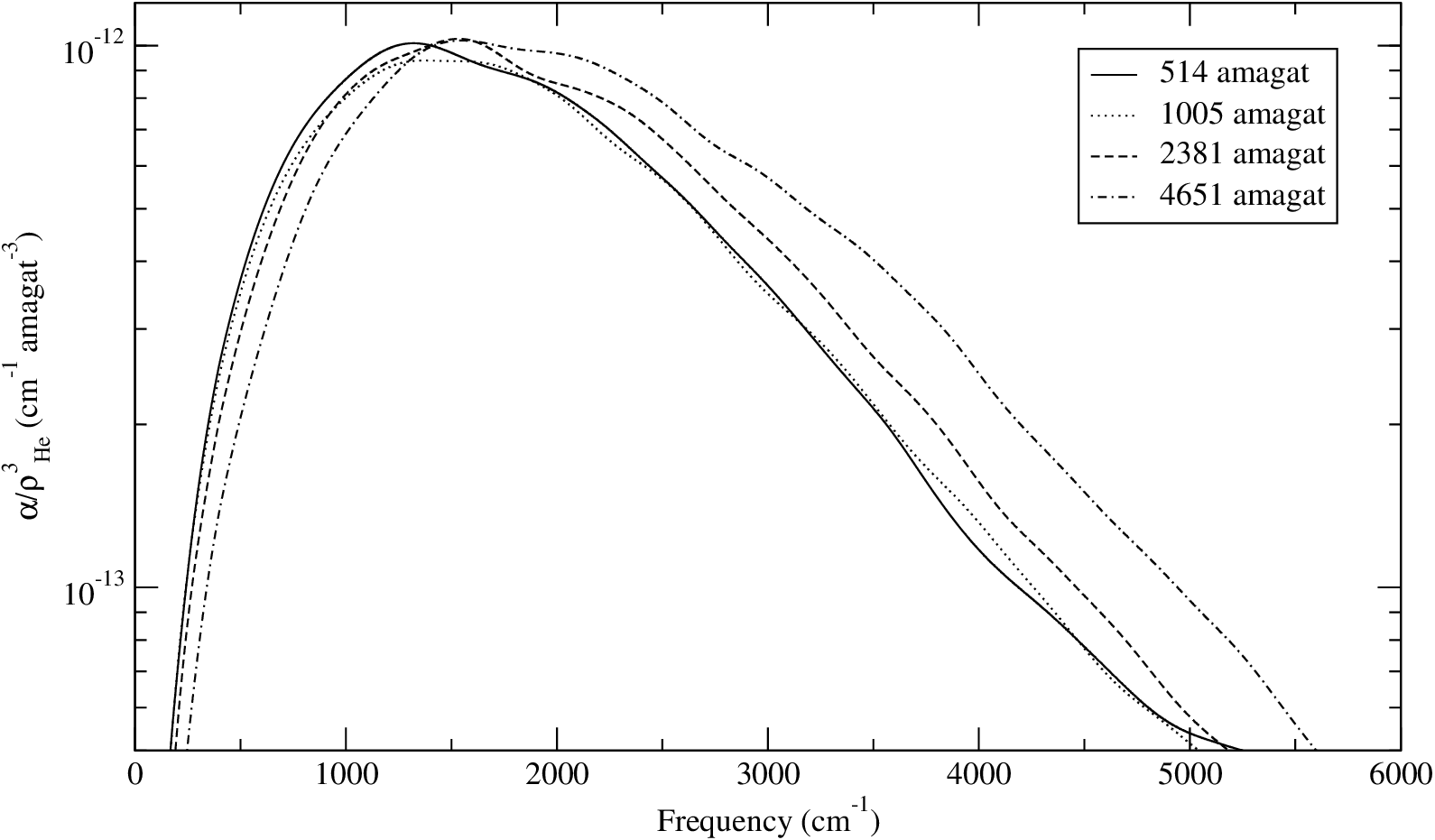}}}
\caption{The simulated IR absorption profiles of dense helium obtained for $T=5000\rm \,K$ and various helium densities indicated in the figure legend. The densities are expressed 
in amagat=$2.68678\cdot 10^{19}\,\rm cm^{-3}$.
\label{F4}}
\end{figure}

\section{Results and discussion}

\subsection{$\rm H-He$ and $\rm H_2-He$ CIA opacities}

To test the applied IR absorption simulation method, we first computed the IR absorption of a supercell containing one H atom or $\rm H_2$ molecule
and 31 He atoms. Such a simulation should result in the reproduction of already known $\rm H-He$ and $\rm H_2-He$ CIA opacities. The comparison of 
the result of our simulations that were performed for $T=\rm 5000\,K$, with the $\rm H-He$ and $\rm H_2-He$ CIA profiles of \citet{GF01} and \citet{AF12} 
is given in Figure \ref{F3}. The simulated IR absorptions reproduce the shapes of both CIA absorption profiles well. The simulations underestimate 
the absorption profiles by a factor of up to $\sim 1.25$, but they correctly predict the magnitude of the absorption strength and the higher frequency 
profiles are also relatively well reproduced.
Overall, this exercise shows that the DFT-based simulations of IR spectra are able to predict correctly the strength and shape of the IR absorption 
of dense $\rm H/He$.

\subsection{CIA opacity of dense helium}

First, we computed the absorption coefficients for temperature of $5000\,\rm K$ and several densities. The result is given in Figure \ref{F4}.
We found that the absorption coefficient is proportional to the cube of density, which is different from the case of $\rm H-He$ and $\rm H_2-He$ CIA opacities,
where the relative absorption coefficients are proportional to the square of density \citep{GF01,AF12}. This is expected because the power of density 
indicates the multiplicity of collisions that contribute to the induction of the dipole moment. During a collision of a pair of identical atoms, 
such as two He atoms, the net dipole moment is zero and such a collision-pair is IR inactive. Therefore, the result given in Figure \ref{F4} shows that the IR absorption 
arises mainly from ternary collisions in dense helium. It is therefore much weaker by about four orders of magnitude than the $\rm H-He$ and $\rm H_2-He$ CIA absorptions 
(comparing results given in Figures \ref{F3} and \ref{F4} and assuming the density of perturbers of about $1000\, \rm amagat$). 
For the most extreme densities represented in Figure \ref{F4} we notice that the absorption profile becomes slightly blueshifted,
which is most probably due to contributions from multiple collisions beyond the ternary ones. Because this effect becomes significant at the most extreme densities 
($>0.8 \,\rm g/cm^3$), at which it should cause just a $400\,\rm \AA$ blueshift of the absorption spectrum, this is a second order effect,
and we neglect it in further analysis. The accurate analysis of the contribution to the IR opacity from more than ternary collisions would require
more extensive studies with larger simulations cells that contain more He atoms, which may be a topic of the subsequent studies.
On the other hand, neglecting this pressure-induced blueshift does not affect the conclusions of the paper.

\begin{figure}[t]
\resizebox{\hsize}{!}{{\includegraphics{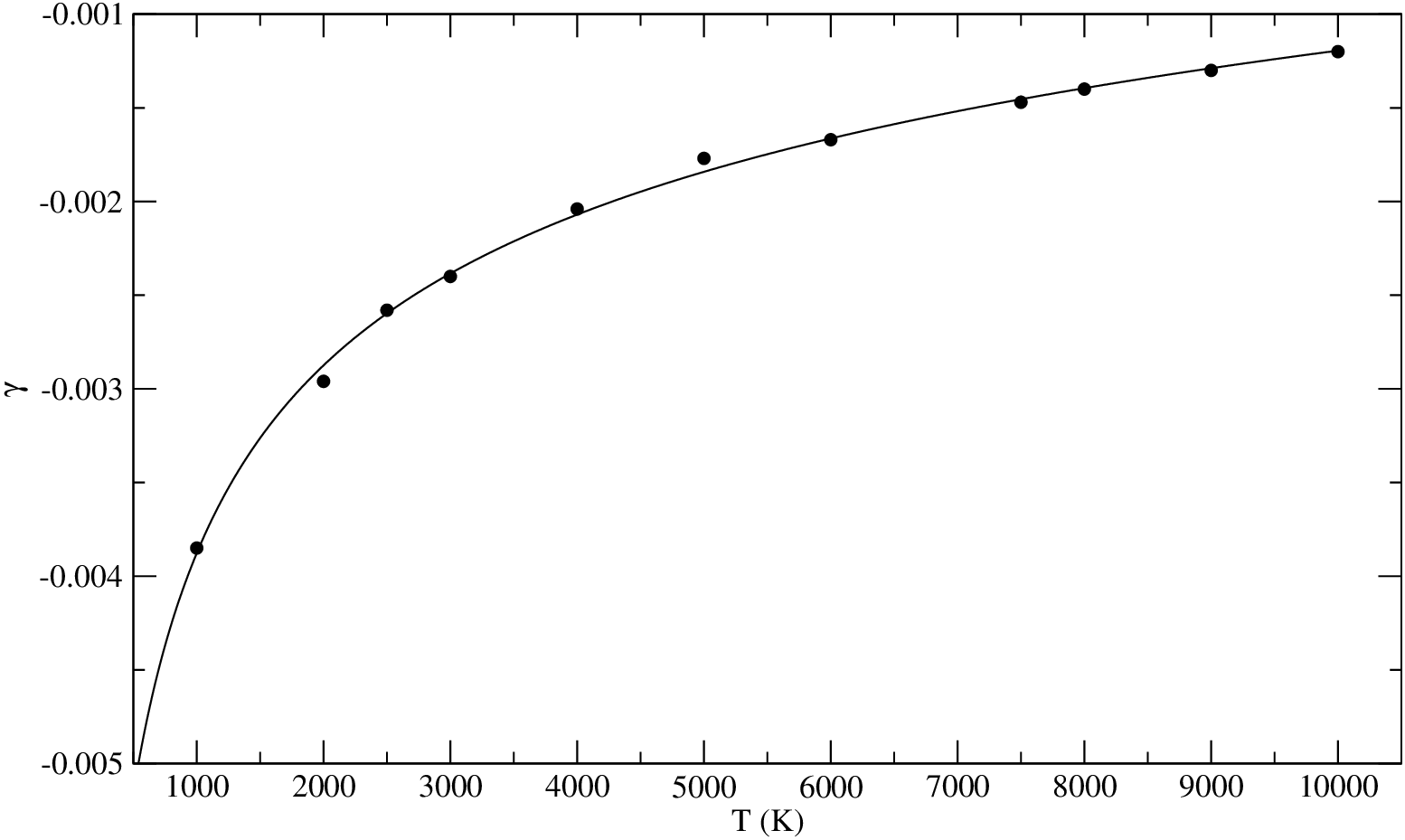}}}
\caption{
The temperature-dependent parameter $\gamma$ as a function of temperature. The points represent the computed values and the line is the best fit given by Eq. \ref{GAM}.
\label{F5}}
\end{figure}

\begin{figure}
\resizebox{\hsize}{!}{{\includegraphics{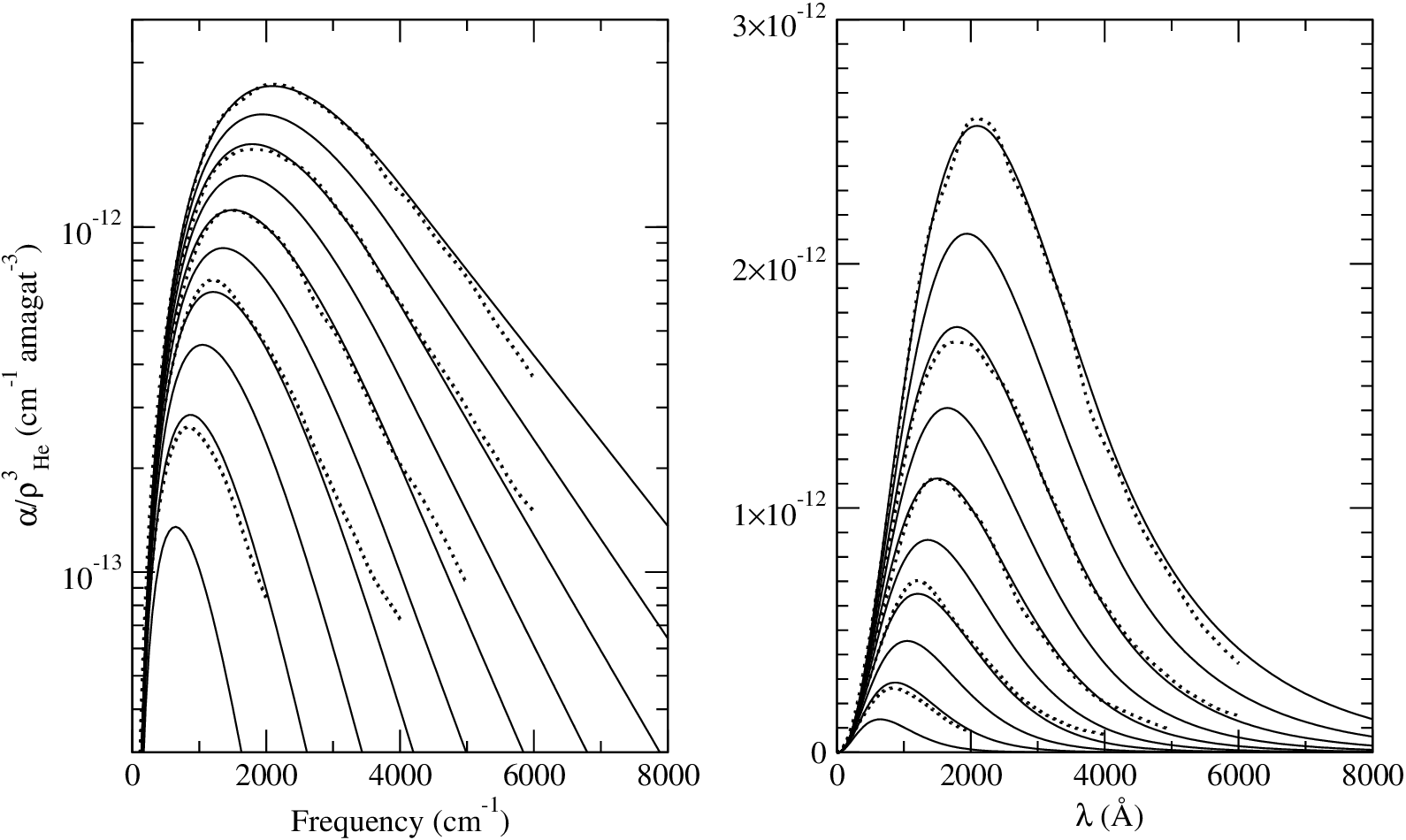}}}
\caption{
The IR opacity of dense helium for different temperatures given by our model (Eqns. \ref{ALP}, \ref{ALP2} and \ref{GAM}.). 
The temperature $T=1000\,\rm K$ to $10000\,\rm K$ with increment of $1000\,\rm K$ (from bottom to top).
The dotted lines represent the results of simulation for $T=2000\,\rm K$, $4000\,\rm K$, $6000\,\rm K$, $8000\,\rm K$, and $10000\,\rm K$. The vertical axis label of right panel 
is identical to that of left panel.
\label{F6}}
\end{figure}

Having found the density dependence of the helium IR absorptivity coefficient we simulated the absorption coefficients for several temperatures of $1000\,\rm K$, $2000\,\rm K$, $2500\,\rm K$, $3000\,\rm K$, $4000\,\rm K$, 
$5000\,\rm K$, $6000\,\rm K$, $7500\,\rm K$, $8000\,\rm K$, $9000\,\rm K$ and $10000\,\rm K$ and fixed helium density of $\rho=514\,\rm amagat=1.38\cdot 10^{22}\,cm^{-3}=0.092\,g/cm^3$.
The resulted absorption profiles were then fitted to the analytical formula:
\begin{equation} \alpha(\omega)/\rho_{\rm He}^3=\beta\omega^{2.5}e^{\gamma(T)\omega}\,{\rm for}\,\,\omega<\omega_{0}\,\,\rm, \label{ALP}\end{equation}
\begin{equation} \alpha(\omega)/\rho_{\rm He}^3=\beta\omega_{0}^{2.5}e^{\gamma(T)\omega_{0}}e^{(\gamma(T)+6.25\cdot10^{-4})(\omega-\omega_{0})} \,\,{\rm for}\,\,\omega>\omega_{0}\,\rm, \label{ALP2}\end{equation}
where $\omega_{0}=4000\rm\,cm^{-1}$ and $\alpha(\omega)/\rho_{\rm He}^3$ is given in $\rm cm^{-1}amagat^{-3}$. The high frequency approximation is constructed, so the value and the derivative over $\omega$ of the two expressions match 
at $\omega=\omega_{0}$ and the $\log (\alpha(\omega)/\rho_{\rm He}^3)$ is a linear function of $\omega$ for large $\omega$.
We selected this simple model to represent the results of the simulations, so it resembles the shape of the simulated absorption and the already known absorption profiles of $\rm H-He$ CIA \citep{GF01}.
The initial fits suggested that the numerical prefactor $\beta$ is temperature independent. Therefore, we decided 
to fix it to the average value of $1.56\cdot10^{-19}$, which was obtained by fitting all the simulated results.
Then we performed one-dimensional fit of $\gamma(T)$ function to the data. The obtained $\gamma(T)$ as a function of temperature is 
\begin{equation} \gamma=(-0.0601248+1.55103\cdot10^{-6}T)\cdot T^{-0.393053} \label{GAM}\end{equation}
and is visualized in Figure \ref{F5} with the simulation results. 
Although complex models of the CIA profiles exist, such as the enhanced Birnbaum-Cohen line shape model used by \citet{GF01},
we found the outlined simple model adequate for our purpose.
The modeled $\alpha(\omega)$ of dense helium is given in Figure \ref{F6}, where we also included the simulated absorption profiles for selected temperatures.
One can see that the overall match of the model to the simulated profiles is pretty good.

\subsection{New synthetic spectra of He-rich white dwarfs}

\begin{figure}
\resizebox{\hsize}{!}{{\includegraphics{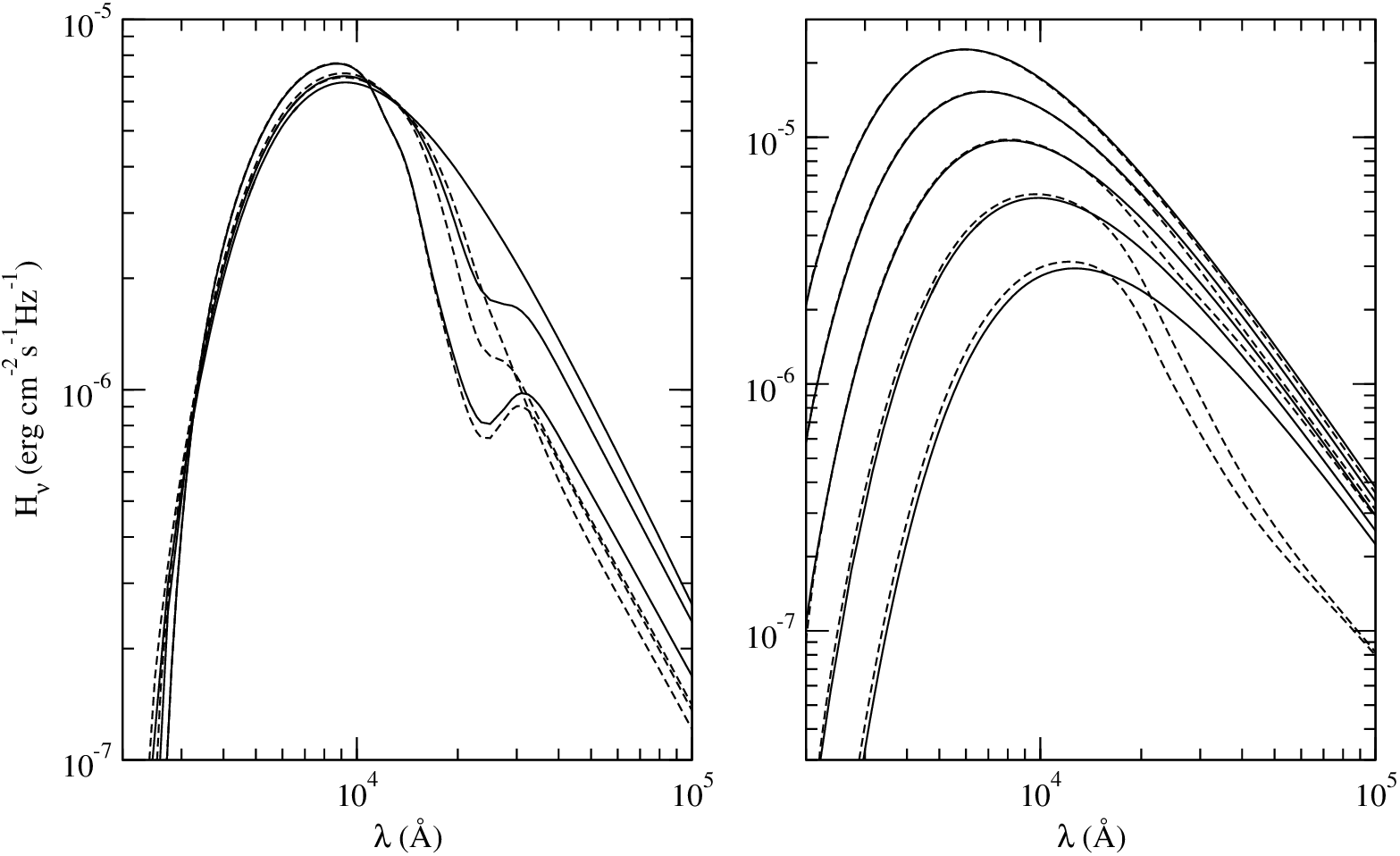}}}
\caption{
Left panel: The synthetic spectra of $T_{\rm eff}=\rm 5300\,K$ and $log\,g=8\,(cgs)$ helium-rich white dwarfs
computed without (solid lines) and with (dashed lines) the IR opacity of helium computed in these studies. 
The three sets of lines represent the models of $\rm He/H=10^{6}$ (resembles the pure He spectrum), $10^{5}$
and $10^{4}$, counting from top to bottom by taking the values at $\lambda=20000\rm \,\AA$. Right panel: 
The synthetic spectra of pure He atmosphere white dwarfs ($log\,g=8\,(cgs)$)
computed without (solid lines) and with (dashed lines) the IR opacity of helium computed in these studies.
The different sets of lines represent the results for $T_{\rm eff}=8000\,\rm K$, $7000\,\rm K$, 
$6000\,\rm K$, $5000\,\rm K$ and $4000\,\rm K$ (from top to bottom). The vertical axis label is identical to that of left panel.
\label{F7}}
\end{figure}

Having the new IR absorption, in the next step we tested its importance in the atmospheres of cool white dwarfs.
In Figure \ref{F7}, we plotted the synthetic spectra of cool, helium-rich stars computed with and without the new opacity.
On the left panel, we plotted the sequence of $T_{\rm eff}=\rm 5300\,K$ and $log\,g=8\,(cgs)$ models. Because of its weaker strength comparing to $\rm H-He$ and $\rm H_2-He$ CIA opacities, 
the new absorption reveals itself in models with $\rm He/H>10^4$. It also significantly reduces the IR fluxes of pure He atmospheres,
as shown in the right panel.
Because the atmosphere becomes more extreme with lowering $T_{\rm eff}$ (see Fig. \ref{F1}), the effect increases with a decrease in $T_{\rm eff}$,
and it reduces the IR flux by $\sim 50\%$ for $T_{\rm eff}\rm=4000\,K$.
On the other hand, it starts to become important only for $T_{\rm eff}\rm <8000\,K$. This is because the atmosphere is significantly less dense  at higher 
$T_{\rm eff}$ and other absorption mechanisms, such as $\rm He^-$ free-free absorption \citep{IRS02}, which strength rises exponentially with temperature, 
become dominant.

In \citet{KSH13}, we have demonstrated that our inability to fit the spectra of He-rich atmosphere stars, such as LHS1126 and LHS3250, 
shows that there may be a problem with the current IR opacities implemented in the atmosphere codes and that the additional IR absorption mechanisms 
may be present in the atmospheres of these stars. Indeed, with the reported CIA opacities of helium, the new synthetic spectra show substantially reduced fluxes 
in IR, and we suspect that the new absorption accounts for at least some of the discrepancy between models and the observed spectra. 
However, the detailed fitting of the spectra of cool white dwarfs requires reliable description of the ionization equilibrium in dense helium, 
which is still rather poorly constrained \citep{KSM07,KSH13}. This is because the ionization fraction determines the strength of the $\rm He^-$ free-free absorption \citep{KSM07},
which interplay with the strength of the IR absorption shapes the spectra of these stars. We therefore avoid such an analysis 
in this study.

\section{Conclusions}

We report a previously unknown IR absorption mechanism resulting from the collisions between He atoms, which we simulated 
by the {\it ab initio} molecular dynamics method.
The new CIA opacity 
is proportional to $\rho_{\rm He}^3$ and arises mainly from the ternary collisions between helium atoms. 
It should dominate the IR absorption in atmospheres of cool, He-rich white dwarfs with $\rm He/H>10^4$, including 
the pure He case, and it may be responsible for problems in fitting 
the spectra of such stars by current models. 
With the discovery of this new absorption mechanism, we are closer to understanding
the absorption processes that prevail in dense helium at extreme conditions, which should result in a better description of the atmospheres 
of cool white dwarfs by models. 

\begin{acknowledgements}
We thank Lothar Frommhold for discussing the issue of collisional induced absorption in dense helium and the manuscript and the reviewer Pierre Bergeron 
as well as Didier Saumon for comments on the manuscript.
\end{acknowledgements}

{}
\end{document}